\documentclass[12pt]{article}
\pdfoutput=1
\usepackage{amsmath,amssymb}
\usepackage{ifpdf}

\ifpdf
  \usepackage[pdftex]{graphicx}
\else
  \usepackage[dvipdfm]{graphicx}
\fi

\numberwithin{equation}{section}

\newcommand{\Ncal}{\mathcal{N}}
\DeclareMathOperator*{\tr}{{\rm tr}}
\newcommand{\del}{\partial}

\newcommand{\Fcal}{\mathcal{F}}
\newcommand{\vol}{\mathrm{vol}}
\newcommand{\vp}{\varphi}

\begin{document}

\thispagestyle{empty}
\begin{flushright}
OU-HET 723
\end{flushright}
\vskip3cm
\begin{center}
{\LARGE {\bf Holographic Interface-Particle Potential}}
\vskip1.5cm
{\large 
Koichi Nagasaki\footnote{nagasaki [at] het.phys.sci.osaka-u.ac.jp}
,\hspace{2mm} Hiroaki Tanida\footnote{hiroaki [at] het.phys.sci.osaka-u.ac.jp}
\,and\hspace{2mm} Satoshi Yamaguchi\footnote{yamaguch [at] het.phys.sci.osaka-u.ac.jp}
}
\vskip.5cm
{\it Department of Physics, Graduate School of Science, 
\\
Osaka University, Toyonaka, Osaka 560-0043, Japan}
\end{center}

\vskip2cm
\begin{abstract}
We consider two $\Ncal=4$ supersymmetric gauge theories connected by an interface and the gravity dual of this system.
This interface is expressed by a fuzzy funnel solution of Nahm's equation in the gauge theory side.  The gravity dual is a probe D5-brane in AdS$_5\times S^5$.  The potential energy between this interface and a test particle is calculated in both the gauge theory side and the gravity side by the expectation value of a Wilson loop.  In the gauge theory it is evaluated by just substituting the classical solution to the Wilson loop. On the other hand it is done by the on-shell action of the fundamental string stretched between the AdS boundary and the D5-brane in the gravity. We show the gauge theory result and the gravity one agree with each other.
\end{abstract}


\newpage
\tableofcontents

\section{Introduction and summary}
A 4-dimensional interface CFT is a scale invariant quantum theory where two same or different 4-dimensional CFTs are connected by a codimension one hyperplane called ``interface.'' This is a kind of generalization of a boundary CFT. Interface CFTs have been discussed in the brane configurations in the string theory \cite{Sethi:1997zza,Ganor:1997jx,Kapustin:1998pb}, and in the AdS/CFT correspondence \cite{Karch:2000gx,DeWolfe:2001pq,Bachas:2001vj} (see also \cite{Kirsch:2004km} and references there in). The supergravity description, sometimes called ``Janus,'' has been successfully constructed \cite{Bak:2003jk,D'Hoker:2006uu,D'Hoker:2006uv,Gomis:2006cu,D'Hoker:2007xy,D'Hoker:2007xz}.  Later the classification and S-duality of the Interface is discussed in \cite{Gaiotto:2008sa,Gaiotto:2008sd,Gaiotto:2008ak}.  More recently 4-dimensional interface CFTs (or boundary CFTs) have been found to be related to knot invariants \cite{Witten:2011zz,Gaiotto:2011nm}.

Let us consider a test particle in the interface CFT.  This test particle, in general, feels a force from the interface. This phenomena is an analogue of the force between dielectric substance and a charged particle in the electromagnetism.

In this paper, we will investigate this phenomena to check the AdS/CFT correspondence \cite{Maldacena:1997re}.  In the gauge theory side we consider the interface described by a fuzzy funnel solution or a ``Nahm pole'' \cite{Constable:1999ac,Gaiotto:2008sa} in $\Ncal=4$ super Yang-Mills theory. This interface connects two $\Ncal=4$ super Yang-Mills theories with the gauge groups SU$(N)$ and SU$(N-k)$.
The potential energy between the interface and the test particle is obtained by evaluating the expectation value of the Wilson loop in the presence of the interface. In this paper, we calculated this potential at the classical level.

In the gravity side, this interface corresponds to a probe D5-brane solution of  \cite{Karch:2000gx} wrapping AdS$_4\times S^2$ subspace in AdS$_5 \times S^5$. On the other hand the Wilson loop corresponds to a fundamental string connecting the AdS boundary and the probe D5-brane.  The potential is obtained by evaluating the on-shell action of this fundamental string \cite{Rey:1998ik,Maldacena:1998im}.

We will find that these two calculation agree with each other.  This is rather surprizing result since the gauge theory calculation is only valid in small 't Hooft coupling $\lambda$ limit and the gravity side calculation is only valid in large $\lambda$ limit.
This agreement is achieved because we have a parameter $k$ which characterize the interface. By virtue of this parameter $k$ we obtain power series of $\lambda/k^2$ in the gravity side; $\lambda/k^2$ is small though $\lambda$ is large in the gravity side. The similar mechanism appears in the BMN limit \cite{Berenstein:2002jq}. This mechanism also appears, for example, in the correlator of a surface operator and a local operator \cite{Drukker:2008wr} (see also \cite{Koh:2008kt}).

The construction of this paper is as follows. In section \ref{sec:gauge} we consider the problem in the gauge theory side. In section \ref{sec:gravity} the calculation in the gravity side is performed and the result is compared with the gauge theory side. In section \ref{sec:general} a generalized problem is discussed. Section \ref{sec:discussion} is devoted to discussions.

\section{Gauge theory side}\label{sec:gauge}
In this section we discuss the setup and calculation in the gauge theory side.
First we give the action and supersymmetries. Next we introduce interface configuration composed of D3- and D5-branes.
Under such circumstances we calculate the potential energy between this interface and a test particle.
This potential has an analogy with Coulomb potential between a dielectric substance and a test charged particle.

\subsection{Action and supersymmetry}
This subsection gives the setup and the action of the $\mathcal{N}=4$ super Yang-Mills theory. 
This theory contains the fields, $A_\mu,\phi_i,\psi : \mu=0,1,\cdots,3,\ i=4,5,\cdots,9$.
These are the gauge field, the real scalar fields and the 16 component spinor, respectively.

The action of this theory is derived from the 10-dimensional super Yang-Mills theory by a trivial dimensional reduction.(See appendix~\ref{Gamma} for the convention of 10-dimensional gamma matrices $\Gamma_M$.)
\begin{align}
 S=\frac{2}{g^2}\int d^4 x \tr\Bigg[
-\frac{1}{4}F_{\mu\nu}F^{\mu\nu}
-\frac{1}{2}D_{\mu}\phi_{i}D^{\mu}\phi_{i}
+\frac{i}{2}\bar{\psi}\Gamma^{\mu}D_{\mu}\psi
+\frac{1}{2}\bar{\psi}\Gamma^{i}[\phi_i,\psi]
+\frac14[\phi_i,\phi_j][\phi_i,\phi_j]
\Bigg],
\end{align}
where the definition of the field strength and the covariant derivative are given by
\begin{align}
& F_{\mu\nu}=\del_{\mu}A_{\nu}-\del_{\nu}A_{\mu}-i[A_{\mu},A_{\nu}],\\
& D_{\mu}\phi_{i}=\del_{\mu}\phi_{i}-i[A_{\mu},\phi_i],\\ 
& D_{\mu}\psi=\del_{\mu}\psi-i[A_{\mu},\psi].
\end{align}
This action possesses the following supersymmetry.
\begin{subequations}
\begin{align}
&\delta A_{\mu}=i\bar{\epsilon}\Gamma_{\mu}\psi,\\
&\delta \phi_{i}=i\bar{\epsilon}\Gamma_{i}\psi,\\
&\delta \psi=\frac12 F_{\mu\nu}\Gamma^{\mu\nu}\epsilon +D_{\mu}\phi_{i}\Gamma^{\mu i}\epsilon
	 -\frac{i}{2}[\phi_{i},\phi_{j}]\Gamma^{ij}\epsilon,
\end{align}
\end{subequations}
where $\epsilon$, the 16 component spinor, is the parameter of the supersymmetry.

\subsection{Interface}
An interface is a codimension one defect which connects two different theories.
Here we consider an interface connecting two $\mathcal{N}=4$ super Yang-Mills theories with gauge groups SU$(N)$ and SU$(N-k)$.
It can be realized in the string theory \cite{Karch:2000gx} as the D3- and D5-brane configuration shown in the table \ref{D3D5system}.

\begin{table}
\begin{center}
\begin{tabular}
{ l || c | c | c | c | c | c | c | c | c | c | c |}
     & 0  & 1  & 2  & 3  & 4 & 5 & 6 & 7 & 8 & 9\\\hline \hline
D3& $\bigcirc$ & $\bigcirc$ &$\bigcirc$& $\bigcirc$ & $\times$   &$\times$ &$\times$ &$\times$  &$\times$ &$\times$\\
D5& $\bigcirc$ & $\bigcirc$ &$\bigcirc$& $\times$ & $\bigcirc$   &$\bigcirc$ &$\bigcirc$ &$\times$  &$\times$ &$\times$\\
\end{tabular}
\end{center}
\caption{D3-D5 system. ``$\bigcirc$'' means the direction the brane is extended, while ``$\times$'' means the normal direction.}\label{D3D5system}
\end{table}

\begin{figure}
\begin{center}
\includegraphics[width=10cm]{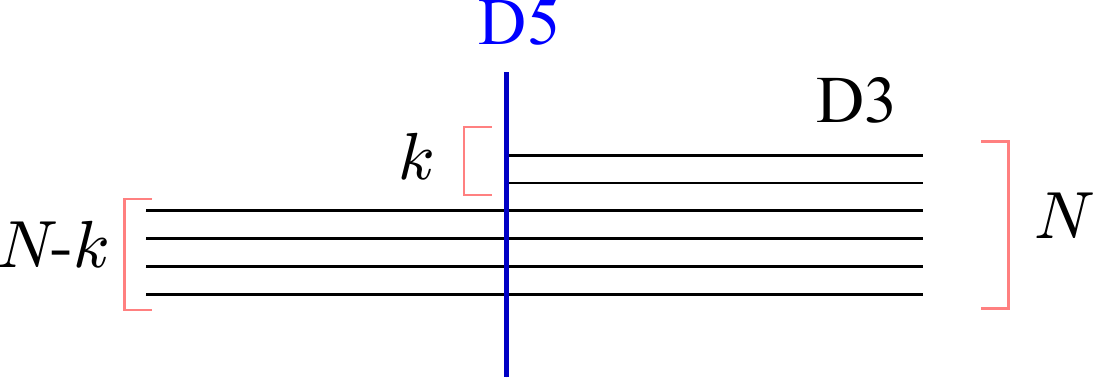}
\end{center}\caption{D3-D5 system. $k$ semi-infinite D3-branes end on a D5-brane.}
\end{figure}

Due to the presence of this interface, the fields have a nontrivial classical vacuum solution.
We analyze the supersymmetry of this classical solution in the gauge theory with the ansatz:
\begin{align*}
 A_{\mu}=0,\qquad \phi_{i}=\phi_{i}(x_3),\ (i=4,5,6),\qquad
\phi_{i}=0,\ (i=7,8,9).
\end{align*}
We obtain the fermion condition
\begin{align}
 0=\delta \psi=\del_{3}\phi_{i}\Gamma^{3i}\epsilon
  -\frac{i}{2}[\phi_i,\phi_j]\Gamma^{ij}\epsilon,
\end{align}
which is rewritten as Nahm's equations:
\begin{align}
 \del_{3}\phi_i=-\frac{i}{2}\epsilon_{ijk}[\phi_j,\phi_k].
\label{Nahm}
\end{align}
The parameters of the remaining supersymmetries satisfy
\begin{align}
 (1-\Gamma^{3456})\epsilon=0.
\end{align}
Nahm's equations \eqref{Nahm} have a fuzzy funnel solution \cite{Constable:1999ac}:
\begin{align}\label{funnel}
 \phi_i=-\frac{1}{x_3}t_i \oplus 0_{(N-k)\times (N-k)},\qquad (x_3>0)
\end{align}
where $t_i,\ i=4,5,6$ are generators of a representation of SU$(2)$. Namely, $t_i$ are $k\times k$ matrices satisfying the commutation relations.
\begin{align*}
& [t_i,t_j]=i\epsilon_{ijk}t_k, \quad i,j,k=4,5,6,\\
& \epsilon_{ijk} \text{: totally anti-symmetric tensor and } \epsilon_{456}=+1.
\end{align*}

In the rest of this paper we only consider $t_i$ of the $k$-dimensional irreducible representation.

\subsection{Test particle and  Wilson loop}
In this subsection we would like to discuss the potential energy between our interface and a test particle.
In order to calculate this potential energy, we adopt the idea of Wilson loop operator $W(z)$ inserted at the distance $z$ from the interface.
It is known that the expectation value of the Wilson loop operator \cite{Wilson:1974sk} is related to the potential energy as 
\begin{equation}\label{WilsonloopPotential}
\langle W(z) \rangle\cong \exp (-TV(z)).
\end{equation}
$T$ denotes the time interval which is taken to be infinity.

Here we introduce this Wilson loop operator and evaluate its expectation value classically.
Let us consider the Wilson loop in Euclidean space.
\begin{align}\label{Wilson}
 W(z)=\tr P \exp \int_{x_3=z} dt (iA_{0}-\phi_4),
\end{align}
where ``$\tr$'' is the trace in the fundamental representation and ``$P$'' means a path-ordered product.
The expectation value of this operator is evaluated classically by substituting the classical solution \eqref{funnel} to eq.~\eqref{Wilson}.
\begin{align}
 \langle W(z) \rangle 
&=\tr P \exp \int dt \left( \frac{1}{z}t_4 \right)\nonumber\\
& =\sum_{\ell:\text{ eigen values of } t_4}\exp \left(T\frac{1}{z}\ell\right)\nonumber\\
& \cong \exp \left(T\frac{1}{z}\ell_{\text{max}}\right) \qquad (T\to \infty)\nonumber\\
& =\exp\left(T\frac{k-1}{2z}\right).
\end{align}
In the last line we used the expression of the maximal eigen value of  $t_4$
\footnote{This is the highest weight of the representation.}, $\ell_\text{max}=\frac{k-1}{2}$.
By using the relation \eqref{WilsonloopPotential} the potential energy in this configuration is
\begin{align}
 V(z)=-\frac{k-1}{2z}.\label{gauge}
\end{align}
We will compare this result with the gravity dual calculation in the next section.

\section{Dual gravity side}\label{sec:gravity}

In this section we compute the potential between interface and a test
particle with the use of the dual gravity description.  We start with an
analysis of a probe D5-brane on the D3-brane background related to the
interface gauge theory following \cite{Karch:2000gx}. The Wilson loop in
the gauge theory corresponds to the classical string
\cite{Rey:1998ik,Maldacena:1998im} which is suspended from the infinite
boundary of background and attached on the probe D5-brane. Finally it is
confirmed that the calculation of the gravity side agrees with the
result of the gauge theory.

\subsection{D3-brane background}

Through the study of AdS/CFT correspondence, it is widely known
that the near horizon geometry of D3-branes, as the solution of
the 10-dimensional type IIB supergravity, equivalently describe the world volume gauge
theory on $N$ D3-branes.
We prepare this gravity background.
The metric takes following AdS$_5\times S^5$ form, using the coordinates
$y,x^{\mu},\ \mu=0,1,2,3$:
\footnote{The boundary
of the AdS$_5$ is at $y=0$.}
\begin{align}
 ds^2=\frac{1}{y^2}(dy^2+dx^{\mu}dx^{\nu}\eta_{\mu\nu})+d\Omega_{5}^2, \label{ads5}
\end{align}
with RR 4-form
\begin{align}
 C_4=-\frac{1}{y^4} dx^{0} dx^{1} dx^{2} dx^{3}+4\alpha_4, \label{rr4}
\end{align}
where $\eta_{\mu\nu}$ and $d\Omega_5^2$ denote 4-dimensional Lorentzian metric
$\eta_{\mu\nu}=\mathrm{diag}(-1,+1,+1,+1)$ and the unit $S^5$ metric, respectively. Here we also use 4-form $\alpha_4$ in $S^5$ which satisfy $d\alpha_4=$(volume form of $S^5$).
In this paper we employ the unit in which the radii of AdS$_5$ and $S^5$ are $1$.
In this unit the slope parameter $\alpha'$ can be written as $\alpha'=1/\sqrt{\lambda}:=1/\sqrt{4\pi g_s N}$, where $g_s$ is the string coupling constant and $\lambda$ corresponds to the 't Hooft coupling in the gauge theory side.

\subsection{Probe D5-brane}

For analyzing the gravity dual to the interface gauge theory, we
put a single probe D5-brane, whose backreaction can be neglected, on the D3-brane background as realization
of the interface.
It is appropriate that we arrange the probe D5-brane on the AdS$_4$ in
the AdS$_5$ and $S^2$ on the equator of the $S^5$. 
The action of a single D5-brane is given by
\begin{align}
 S_{D5}=-T_5 \int \sqrt{-\det (G+\Fcal)}+T_5\int \Fcal C_4,
\end{align}
which consists of two terms; the first term is the Dirac-Born-Infeld action and the second term is the Wess-Zumino term.
Here we set the pull-back of metric as $G$ and world volume gauge flux as $\Fcal$.
And the D$p$-brane tension is defined as
\begin{align}
 T_{p}=\frac{1}{(2\pi)^{p}\alpha'^{(p+1)/2}g_s}.
\end{align}

Consider the solution of probe D5-brane in the background (\ref{ads5}) and (\ref{rr4}) under the following ansatz
\begin{align}
 y=y(x_3),\qquad \Fcal=-\kappa \;\vol [S^2],
\end{align}
with a constant $\kappa$ and the $S^2$ volume form $\mathrm{vol}[S^2]$.
By substituting the ansatz, we can rewrite the action
\begin{align}
 S_{D5}=-4\pi T_5 V \int
 dx_3\frac{1}{y^4}\left(\sqrt{\left(\left(\partial_3 y\right)^2+1\right)(1+\kappa^2)}-\kappa\right),
\end{align}
where $V$ means volume of the 3-dimensional subspace along
 $(x^0,x^1,x^2)$ directions in the probe D5-brane and we use
 $\partial_3$ instead of $\partial/\partial x_3$.
We solve the equation of motion
\begin{align}
 \partial_3\left(\frac{\partial_3y}{y^4}\sqrt{\frac{1+\kappa^2}{\left(\partial_3y\right)^2+1}}\right)
+\frac{4}{y^5}\left(\sqrt{\left(\left(\partial_3y\right)^2+1\right)(1+\kappa^2)}-\kappa\right)=0,
\end{align}
and obtain the solution of probe D5-brane
\begin{align}
 x_3=\kappa y, \label{d5sol}
\end{align}
which fixes the position of probe D5-brane located on the AdS$_5$.
In addition, charges of D3-branes appear as magnetic flux in the D5-brane
world volume, because the D5-brane are linked to D3-branes through the fuzzy
funnel solution \eqref{funnel} in the world volume theory. Namely we can associate $k$ with $\kappa$
\begin{align}
 k&=-\frac{T_5}{T_3}\int\Fcal=\frac{\kappa}{\pi\alpha'}.
\end{align}

\subsection{String and potential}
\label{sec:string}
Now let us focus a string ending on the probe D5-brane from the infinite
distance corresponds to the Wilson loop \eqref{Wilson} in the
gauge theory with interface. Therefore we can identify the
interface-particle potential from the on-shell string action.

\begin{figure}
\begin{center}
\includegraphics[width=10cm]{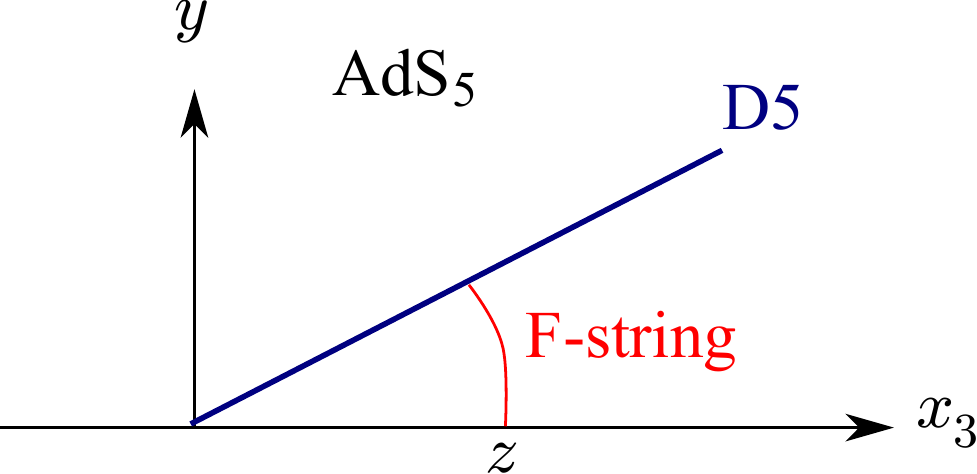}
\end{center}\caption{The probe D5-brane and the fundamental string in the AdS$_5$ expressed by the solution \eqref{d5sol}. }
\end{figure}

In the conformal gauge, the Polyakov action and the Virasoro constraints are
\begin{align}
 &S=\frac{1}{4\pi \alpha'}\int d \tau d \sigma(\dot{X}^{M}\dot{X}_{M}+X'^{M}X'_{M}),\\
 &\dot{X}^{M}\dot{X}_{M}-X'^{M}X'_{M}=0,\qquad \dot{X}^{M}X'_{M}=0,
\end{align}
where $\tau, \sigma$ are string world sheet coordinates and differentials
with respect to them are denoted by ``$\,\,\dot{}\,\,$''
and ``$\,\,'\,\,$'' respectively. We assume the region of $\sigma$ as $0\le \sigma \le \sigma_1$. The string ends on the AdS boundary at $\sigma=0$ and is attached to the D5-brane at $\sigma=\sigma_1$.
We can set following ansatz for the string to be static:
\begin{align}
 t=t(\tau),\quad y=y(\sigma),\quad x_3=x_3(\sigma).
\end{align}
Then the action and the constraints are translated into
\begin{align}
& S=\frac{1}{4\pi \alpha'}\int d \tau d \sigma \frac{1}{y^2}(\dot{t}^2+y'^2+x_3'^2),\\
& \dot{t}^2=y'^2+x_3'^2.
\end{align}
The equations of motion are given by
\begin{align}
&\ddot{t}=0, \label{eqm1}\\
&\left(\frac{x_3'}{y^2}\right)'=0, \label{eqm2}\\
&-\frac{2}{y^3} (\dot{t}^2+y'^2+x_3'^2)-\left(\frac{2y'}{y^2}\right)'=0. \label{eqm3}
\end{align}
Note that we impose boundary conditions
\begin{align}
\kappa x_3'(\sigma_1)+y'(\sigma_1)=0, \label{bc1}\\
-x_3(\sigma_1)+\kappa y(\sigma_1)=0. \label{bc2}
\end{align}
These boundary conditions denote the string is attached to the
probe D5-brane.
In particular the first line is Neumann boundary condition along the probe
D5-brane and the second line is Dirichlet boundary condition transverse to the probe D5-brane.

Next we solve the equations of motion with above boundary conditions
under the gauge $t=\tau$.
Eq.~\eqref{eqm2} gives
\begin{align}
 \frac{x_3'}{y^2}=-c,\qquad \left(c\text{ : constant}\right).\label{x3'}
\end{align}
And the Virasoro constraint becomes
\begin{align}
 -1+y'^2+c^2 y^4=0,
\end{align}
which takes the form
\begin{align}
 y'=\sqrt{1-c^2 y^4}.\label{y'}
\end{align}
Taking the boundary condition $x_3(0)=z$ into account, eq.~\eqref{x3'} is solved as
\begin{align}
 \int_{z}^{x_3}dx_3&=-c \int_0^{\sigma}d\sigma y^2
 =-c\int_0^{y}dy \frac{y^2}{y'}
 =-c\int_0^{y}dy \frac{y^2}{\sqrt{1-c^2y^4}}\notag\\
 x_3-z&=-\frac{1}{\sqrt{c}}(E(\vp,i)-F(\vp,i)), \label{x3}
\end{align}
where we introduced the elliptic integrals $E(\vp,i)$ and $F(\vp,i)$  for convenience
(see appendix \ref{app:elliptic} for detail).
The boundary condition (\ref{bc1}) indicate $y_1=y(\sigma_1)$ by using (\ref{x3'}) and (\ref{y'}),
\begin{align}
 \sqrt{c}y_1=(1+\kappa^2)^{-1/4}.
\end{align}
On the other hand, we can solve the boundary condition (\ref{bc2})
and determine the constant $c$
\begin{align}
 \sqrt{c}=\frac{1}{z}[E(\vp_1,i)-F(\vp_1,i)+\kappa(1+\kappa^2)^{-1/4}],\qquad
 \left(\sin \vp_1 :=\sqrt{c}y_1\right).
\end{align}

With the use of the formula
\begin{align}
 &\frac{1}{u^2\sqrt{(1-u^2)(1-h^2u^2)}}\notag\\
&=\frac{d}{du}\left[
-\frac{1}{u}\sqrt{(1-u^2)(1-h^2u^2)}
\right]
-\sqrt{\frac{1-h^2u^2}{1-u^2}}+\frac{1}{\sqrt{(1-u^2)(1-h^2u^2)}},
\end{align}
we can rewrite the action
\begin{align}
 S&=\frac{1}{4\pi \alpha'}\int d\tau d\sigma
\frac{1}{y^2}\left(
\dot{t}^2+y'^2+x_3'^2
\right)\nonumber\\
&=\frac{T}{4\pi\alpha'} \int_{\epsilon}^{\sigma_1}d\sigma \frac{2}{y^2}\nonumber\\
&=\frac{T}{2\pi\alpha'} \int_{\epsilon}^{y_1}dy \frac{1}{y^2\sqrt{1-c^2y^4}}\nonumber\\
&=\frac{T}{2\pi\alpha'} \sqrt{c} \left(
\frac{1}{\sqrt{c}\epsilon}+O(\epsilon)
-\frac{\sqrt{1-c^2 y_1^4}}{\sqrt{c}y_1}-E(\vp_1,i)+F(\vp_1,i)
\right),\label{s}
\end{align}
where we chose the integral region $\epsilon\rightarrow y_1$, due to decompose the divergence
originate with the string self-energy.
The potential piece, to compare with the gauge theory, is extracted by removing the divergence from (\ref{s}) as in \cite{Rey:1998ik,Maldacena:1998im,Drukker:1999zq}.
The potential is read off from \eqref{s} as
\begin{align}
V(z)=&\frac{1}{2\pi\alpha'} \sqrt{c} \left(
-\frac{\sqrt{1-c^2 y_1^4}}{\sqrt{c}y_1}-E(\vp_1,i)+F(\vp_1,i)
\right)\nonumber\\
=&-\frac{1}{2\pi\alpha'z}\left(
\frac{\kappa}{(1+\kappa^2)^{1/4}}
+E(\vp_1,i)-F(\vp_1,i)\right)^2,
\label{gravity-potential}
\end{align}
where $\vp_1$ is defined as $\sin\vp_1=(1+\kappa^2)^{-1/4}$.

\subsection{Large $\kappa$ limit}
We conclude this section by estimating the potential with large $\kappa=\pi k/\sqrt{\lambda}$
limit. Since we assume $N\gg \kappa$, large $\kappa$ limit does not
affect to the gravity background. 
Then the potential \eqref{gravity-potential} is expanded as
\begin{align}
 V=-\frac{k}{2z}\left(1+\frac{1}{6\pi^2}\frac{\lambda}{k^2}+O(\frac{\lambda^2}{k^4})\right).
\end{align}
Even if $\lambda$ is large, $\lambda/k^2$ can be small. Thus this expansion is formally positive power series of $\lambda$ and could be compared with the gauge theory side.  At the leading contribution, we confirmed the AdS/CFT correspondence of the interface-particle potential \eqref{gauge} in the gauge theory picture.  The next to leading term is the prediction for the $\lambda$ correction in the gauge theory side.

\section{Generalization}\label{sec:general}
In this section we consider a kind of generalization for the test particle while the interface is not changed. We compute the potential energy between the interface and this generalized test particle both in the gauge theory side and the gravity side. Those two results agree to each other in the leading order.

\subsection{Gauge theory side}
We consider a test particle parameterized by $\chi$, $0\le \chi \le \pi/2$ expressed by the Wilson loop
\begin{align}
 W(z,\chi)=\tr P \exp \int dt(iA_{0}-\sin\chi \phi_4-\cos \chi \phi_7).
\label{Wilson2}
\end{align}
When $\chi=\pi/2$ this particle is the same as the previous one, while this is mutually supersymmetric to the interface when $\chi=0$.

The potential energy between the interface and this generalized test particle is evaluated by substituting the solution \eqref{funnel}\footnote{$\phi_7=0$ in this solution} to the Wilson loop \eqref{Wilson2} as the same way as before. The result turns out to be
\begin{align}
 V(z)=-\frac{k-1}{2z} \sin\chi.\label{gauge2}
\end{align}
This is the same as eq.~\eqref{gauge} when $\chi=\pi/2$. On the other hand the potential energy \eqref{gauge2} vanishes when $\chi=0$ as expected since the interface and the test particle are mutually supersymmetric in this case.

\subsection{Gravity side}\label{sec:generalized-gravity}

\begin{figure}
 \begin{center}
  \includegraphics{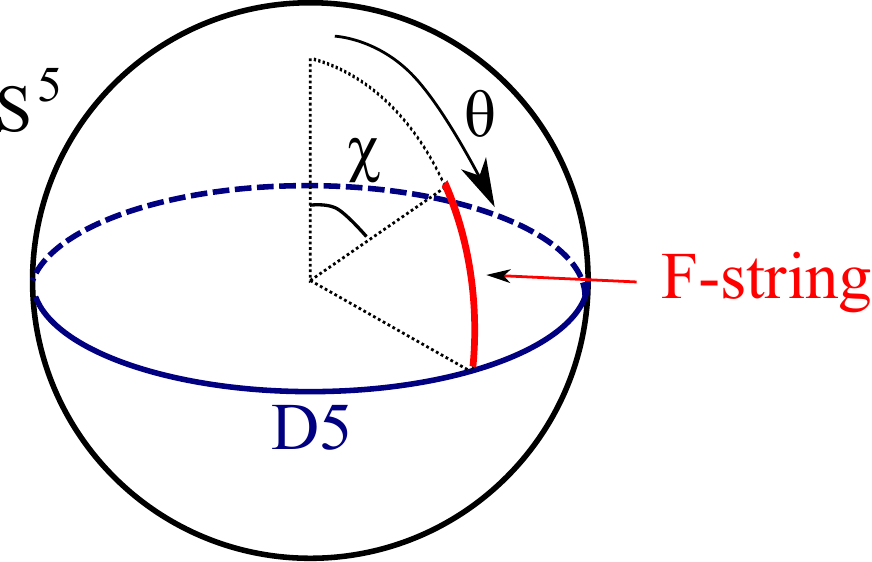}
 \end{center}
\caption{The D5-brane and F-string configuration on the $S^5$.}
\label{fig:s}
\end{figure}

Here in this subsection, we calculate the potential between the interface and the generalized test particle in the gravity side as the same way as section \ref{sec:string}.
The only difference is the boundary condition at $y=0$. Let $\theta$ be the angle from the North pole of $S^5$ as shown in figure \ref{fig:s}. We impose the boundary condition $\theta=\chi$ at $y=0$ and $\theta=\pi/2$ at the other end of the string.

Here we just show the result. See appendix \ref{app:calculation} for the detail of the calculation. 
Eqs.\eqref{bcd0},\eqref{bcd1},\eqref{bcd2} give three equations for three unknowns $y_1, m,c$.
\begin{align}
& \frac{\pi}{2}-\chi=\frac{m}{\sqrt{A}}F(\vp_1,h),\\
& 1-m^2y_1^2-c^2(1+\kappa^2)y_1^4=0,\\
& \kappa y_1=z+\frac{c}{\sqrt{A}B}(E(\vp_1,h)-F(\vp_1,h)),
\end{align}
where we use the notation for short hand:
\begin{equation}
\begin{aligned}
 &A:=\frac12(m^2+\sqrt{m^4+4c^2}),\quad B:=\frac12(m^2-\sqrt{m^4+4c^2}),\\
 &h^2:=\frac{B}{A},\quad \sin\vp_1:=\frac{y_1}{\sqrt{A}} .
\end{aligned}
\end{equation}
The potential is written as
\begin{align}
 V(z)=\frac{1}{2\pi\alpha'}
\sqrt{A}\left[-\frac{\cos\vp_1}{\sin\vp_1}\sqrt{(1-\frac{B}{A} \sin^2\vp_1)}-E(\vp_1,h)+F(\vp_1,h) \right],
\end{align}
As in the previous case we can estimate this potential in the limit $\kappa \to \infty$ as
\begin{align}
V(z)=-\frac{k\sin\chi}{2z}\left[1
+\frac{\sin\chi}{4\kappa^2\cos^3\chi}\left(
\frac{\pi}{2}-\chi-\frac12\sin2\chi
\right)
+O(\kappa^{-4})\right].
\end{align}
The leading term in this expansion agrees with the gauge theory side
\eqref{gauge2} (if $k$ is large) and the second term gives the
prediction for $\kappa^{-2}=\frac{\lambda}{\pi^2 k^2}$ correction.

\section{Discussion}\label{sec:discussion}
In this paper we investigate the 1/2 BPS interface, in particular the
potential between this interface and a test particle.  We calculated the
potential both in the gauge theory side and the gravity side and found
perfect agreement in the leading order. This is a strong evidence of the
AdS/CFT correspondence including the interface.

In the gravity side we also obtained sub-leading corrections of a power
series of $\lambda/k^2$.  This may be compared to the perturbative
corrections in the gauge theory side.  It will be an interesting future work to
calculate these sub-leading corrections in the gauge theory side and see
if they agree with the gravity side.

Here we give a heuristic argument on the perturbative corrections in the
gauge theory side, in particular the $\lambda/k^2$ behavior of the
corrections.  This argument is the similar one as in \cite{Rey-Yamaguchi}
\footnote{S.Y. would like to thank Soo-Jong Rey for the discussion.}. In
order to calculate the perturbative corrections, we express the field as
$\phi_i=\phi_i^{(0)}+\tilde{\phi}_i$ where $\phi_{i}^{(0)},\ (i=4,5,6)$
are the classical solution \eqref{funnel} and $\tilde{\phi}_i$ are the
fluctuations of the fields. For simplicity let us perform the following Weyl
transformation and go to $AdS_4$ frame
\begin{align}
 A_{\mu}\to A_{\mu},\quad \psi\to e^{3\Omega/2}\psi, \quad \phi_i \to
 e^{\Omega}\phi_i,
\quad (e^{\Omega}:=r/x_3),
\end{align}
where $r$ is a constant.
The metric becomes by this Weyl transformation
\begin{align}
 ds^2=\frac{r^2}{x_3^2}(\eta_{\mu\nu}dx^{\mu}dx^{\nu}).
\end{align}
The classical solution of \eqref{funnel} is now simply the constant
vacuum expectation value
\begin{align}
 \phi_i^{(0)}=\frac{1}{r} t^{i}\oplus 0_{(N-k)\times (N-k)}, \quad (i=4,5,6).
\end{align}
This is an analogue as the Higgs mechanism. Actually the gauge
field get mass as the following way. The Lagrangian density includes
\begin{align}
 \tr([A_{\mu},\phi_i^{(0)}]^2)=-\frac{1}{2r^2}k^2\tr_{k\times k}(A_{\mu}A_{\mu})+\cdots.
\end{align}
Some of the scalar fields also have mass square term proportional to
$k^2$. These $k^2$ terms are the leading terms in the Lagrangian density
in the large $k$ limit.  Therefore the action can be written as
\begin{align}
 S=N\frac{k^2}{\lambda}\int d^4x\sqrt{g}\mathcal{L}',\quad 
\end{align}
where $\mathcal{L}'$ is a function of the fields and their derivatives, which satisfies
\begin{align}
 \lim_{k\to \infty} \mathcal{L}'=(\text{finite}).
\end{align}
From this form of the action we expect that the perturbative corrections
will be a power series of $\lambda/k^2$ in the large $k$ limit.

\subsection*{Acknowledgments}
We would like to thank Tohru Eguchi, Yosuke Imamura, Shinsuke Kawai,
Teruhiko Kawano, Sanefumi Moriyama, Takahiro Nishinaka, Soo-Jong Rey,
and Tadashi Okazaki for discussions and comments.  S.Y. was supported in
part by KAKENHI 22740165.

\appendix
\section{Gamma matrices}\label{Gamma}
$\Gamma_M,\ M=0,1,\cdots, 9$ are the 10-dimensional gamma matrices satisfying the algebra
\begin{equation}
\{\Gamma_M,\Gamma_N\}=2\eta_{MN}.
\end{equation}
$\eta_{MN}=\mathrm{diag}(-1,+1,\cdots,+1)$ is the metric of 10-dimensional Minkowski space. 
We also use the matrices with anti-symmetric indices.
\begin{equation}
 \Gamma_{MN}=\frac{1}{2}(\Gamma_M\Gamma_N-\Gamma_N\Gamma_M).
\end{equation}

\section{Elliptic integrals}
\label{app:elliptic}
The definition of elliptic integrals
 are
\begin{align}
 &F(\varphi,h):= \int_0^{\sin \varphi}
 \frac{du}{\sqrt{(1-u^2)(1-h^2 u^2)}} &&\text{: the first kind,}\\
&E(\varphi,h):= \int_0^{\sin \varphi}du\sqrt{\frac{1-h^2u^2}{1-u^2}} &&\text{: the second kind.}
\end{align}
And we give a useful formula
\begin{align}
 F(\varphi,h)-E(\varphi,h)=\int_0^{\sin \varphi}du\frac{h^2u^2}{\sqrt{(1-u^2)(1-h^2u^2)}}.
\end{align}

\section{Detailed calculation in the gravity side of the generalized case}
\label{app:calculation}
In this appendix we show the detailed calculation of the potential discussed in section \ref{sec:generalized-gravity}.

Let us put the ansatz:
\begin{align}
 t=t(\tau),\quad y=y(\sigma),\quad x_3=x_3(\sigma),\quad \theta=\theta(\sigma).
\end{align}
Then the action becomes
\begin{align}
 S=\frac{1}{4\pi\alpha'}\int d\tau d\sigma\left[ \frac{1}{y^2}(\dot{t}^2+y'^2+x_3'^2)+\theta'^2\right].
\end{align}
$t=\tau$ is a solution of the equation of motion for $t$.
The equation of motion for $\theta$ is simply  $\theta''=0$. This can be integrated as
\begin{align}
 \theta'=m=(\text{constant}).
\end{align}
$x_3$ is solved just the same way as \eqref{x3'} and thus the
Virasoro constraint becomes
\begin{align}
 \frac{1}{y^2}(-1+y'^2+c^2 y^4)+m^2=0,
\end{align}
and the expression of $y'$ as
\begin{align}
 y'=\sqrt{1-m^2y^2-c^2y^4}.\label{y1g}
\end{align}
Integration of this equation gives the relation between $\sigma_1$(upper bound for $\sigma$) and $y_1:=y(\sigma_1)$ as
\begin{align}
 \int_0^{y_1}dy\frac{1}{\sqrt{1-m^2y^2-c^2y^4}}=\sigma_1.
\end{align}
It is convenient to introduce the number $A,B$:
\begin{align}
 &A=\frac12(m^2+\sqrt{m^4+4c^2}),\\
 &B=\frac12(m^2-\sqrt{m^4+4c^2}),
\end{align}
since we can rewrite the inside the square root in eq.~\eqref{y1g} as
\begin{align}
 1-m^2y^2-c^2y^4=(1-Ay^2)(1-By^2).
\end{align}
Notice that $B<0<A$ is satisfied.
Eq.~\eqref{x3'} can be integrated and gives the value of $x_3$ at $\sigma=\sigma_1$.
\begin{align}
 x_3(\sigma_1)=z+cA^{-1/2}\frac{1}{B}(E(\vp_1,h)-F(\vp_1,h)),\label{x3g}
\end{align}
where $\sin\vp_1=\sqrt{A}y_1$.
$\theta$ is also solved as
\begin{align}
 \theta=m\sigma+\chi.
\end{align}
Since $\theta(0)=\chi$ and $\theta(\sigma_1)=\frac{\pi}{2}$ we obtain
\begin{align}
 \frac{\pi}{2}-\chi=m\sigma_1.\label{bcd0}
\end{align}
At $\sigma=\sigma_1$ we should impose the boundary conditions. One of them is
\begin{align}
 \kappa x'_3(\sigma_1)+y'(\sigma_1)=0.
\end{align}
This equation can be rewritten as
\begin{align}
 1-m^2y_1^2-c^2(1+\kappa^2)y_1^4=0,\label{bcd1}
\end{align}
where we use eq.~\eqref{y1g} and eq.~\eqref{x3'}. 
The other boundary condition at $\sigma=\sigma_1$:
\begin{align}
 -x_3(\sigma_1)+\kappa y_1=0.
\end{align}
Substituting $x_3(\sigma_1)$ by \eqref{x3g} we obtain
\begin{align}
 \kappa y_1=z+\frac{c}{\sqrt{A}B}(E(\vp_1,h)-F(\vp_1,h)).\label{bcd2}
\end{align}

The action becomes
\begin{align}
 S&=\frac{T}{2\pi \alpha'}
\sqrt{A}\left[\frac{1}{\sqrt{A}\epsilon}+O(\epsilon)-\frac{\cos\vp_1}{\sin\vp_1}\sqrt{(1-\frac{B}{A} \sin^2\vp_1)}-E(\vp_1,h)+F(\vp_1,h) \right].
\end{align}
Thus the regularized action $S_{\mathrm{reg}}$ is obtained by subtracting the divergent part. We can then read off the potential from $S_{\mathrm{reg}}$ as
\begin{align}
 V(z)=\frac{1}{2\pi\alpha'}
\sqrt{A}\left[-\frac{\cos\vp_1}{\sin\vp_1}\sqrt{(1-\frac{B}{A} \sin^2\vp_1)}-E(\vp_1,h)+F(\vp_1,h) \right].
\end{align}

\providecommand{\href}[2]{#2}\begingroup\raggedright\endgroup

\end{document}